%% file: aaai25.tex
\def\year{2025}\relax

\documentclass[letterpaper]{article} % DO NOT CHANGE THIS
\usepackage{aaai25}  % DO NOT CHANGE THIS
\usepackage{times}  % DO NOT CHANGE THIS
\usepackage{helvet}  % DO NOT CHANGE THIS
\usepackage{courier}  % DO NOT CHANGE THIS
\usepackage[hyphens]{url}  % DO NOT CHANGE THIS
\usepackage{graphicx} % DO NOT CHANGE THIS
\usepackage{natbib}  % DO NOT CHANGE THIS AND DO NOT ADD ANY OPTIONS TO IT
\usepackage{caption} % DO NOT CHANGE THIS AND DO NOT ADD ANY OPTIONS TO IT
\usepackage{booktabs}
\usepackage{enumitem}
\usepackage{multirow}
\DeclareCaptionStyle{ruled}{labelfont=normalfont,labelsep=colon,strut=off} % DO NOT CHANGE THIS
\usepackage{algorithm}
\usepackage{algorithmic}
\usepackage{makecell}

\usepackage{xcolor}

\newcommand{\appx}{\textit{Appendices}}
\newcommand{\sectionname}[2]{\expandafter\gdef\csname #1\endcsname{#2}}

\newcommand{\secref}[1]{\textit{\csname #1\endcsname}}

\usepackage{newfloat}
\usepackage{listings}
\floatstyle{ruled}
\newfloat{listing}{tb}{lst}{}
\floatname{listing}{Listing}

\title{Identifying Constructive Conflict in Online Discussions \\through Controversial yet Toxicity Resilient Posts}
\author{
    Ozgur Can Seckin\thanks{Corresponding author. Email: oseckin@iu.edu \\$^\dag$Equal contributions.}$^\dag$, Bao Tran Truong$^\dag$, Alessandro Flammini, Filippo Menczer}
\affiliations{
Observatory on Social Media, Indiana University, Bloomington
}

\begin{document}

\maketitle

\begin{abstract}

Bridging content that brings together individuals with opposing viewpoints on social media remains elusive, overshadowed by echo chambers and toxic exchanges.
We propose that algorithmic curation could surface such content by considering \emph{constructive conflicts} as a foundational criterion. We operationalize this criterion through \emph{controversiality} to identify challenging dialogues and \emph{toxicity resilience} to capture respectful conversations. 
We develop high-accuracy models to capture these dimensions.
Analyses based on these models demonstrate that assessing resilience to toxic responses is not the same as identifying low-toxicity posts. 
We also find that political posts are often controversial and tend to attract more toxic responses.
However, some posts, even the political ones, are resilient to toxicity despite being highly controversial, potentially sparking civil engagement.
Toxicity resilient posts tend to use politeness cues, such as showing gratitude and hedging.
These findings suggest the potential for framing the tone of posts to encourage constructive political discussions.
 
\end{abstract}

\section{Introduction}

Polarization has intensified globally in recent years \cite{boxell2024cross}, driven in part by the growing influence of social media \cite{lorenz2023systematic}. In contrast to the original promise of social media as a modern public square that fosters discussions, echo chambers emerge where users preferentially consume content aligned with their ideologies~\cite{garimella2017long,flaxman2016filter, barbera2015tweeting, conover2011political}. Algorithms further amplify this selective exposure, segregating users into even more polarized groups, exacerbating affective polarization~\cite{santos2021link, cho2020search, sasahara2021social}.

Theoretically, cross-cutting discussions can promote mutual understanding and greater tolerance for diversity \cite{kingwell1994civil, gutmann2009democracy, bohman1997deliberative}, as well as correcting misperceptions about political out-groups \cite{voelkel2023interventions}. As such, encouraging individuals to engage with diverse perspectives can be a potential solution to the selective exposure problem and could reduce affective polarization.

However, empirical evidence shows that exposure to opposing views can sometimes reinforce existing beliefs, intensifying polarization~\cite{bail2018exposure}.
One of the possible causes is the negative tone accompanying diverse exposure~\cite{efstratiou2023non}, which can emphasize ingrained negative emotions toward outgroups \cite{lerman2024affective}.
This could lead to hostility rather than fostering understanding.
The impact of diverse exposure also hinges on the topic. For instance, positive outcomes are less likely in conversations about contentious topics \cite{santoro2022promise}.
Together, the tone and topic of conversations might explain the mixed empirical findings on the effects of diverse exposure on social media \cite{bail2018exposure,levy2021social, santoro2022promise}.

To effectively bridge people across divides, interventions must consider the nuance in both the tone and topic of conversations. The above literature suggests that interventions introducing diverse exposure hold promise only when disagreements remain civil. However, across platforms, controversies are consistently associated with increased toxicity \cite{avalle2024persistent}.
This raises key questions: Can conversations about critical democratic issues be respectful? Are controversial political topics inherently prone to hostility?

We hypothesize that even highly controversial political topics can be framed in ways that reduce toxic responses --- this is supported by the literature on conversational framing \cite{bao2021alrightconversations}. Qualitative work has shown that toxicity is not intrinsic to political content. For example, topics often associated with high toxicity, such as war and conflicts, are not linked with toxicity when approached from a humanistic perspective. Prior work also demonstrates that toxicity is not tied to specific user groups \cite{mall2020four}, suggesting that politically active users can engage in civil discourse. 
Understanding the interplay between tone and topic, especially in topics where people tend to disagree, is critical for designing appropriate interventions. 

While prior research has examined online controversies \cite{garimella2018quantifying, mejova_controversy_2014} and the influence of conversational tone on prosociality \cite{bao2021alrightconversations} or incivility \cite{zhang2018awryconversations, almerekhi2019toxicitytriggers}, no work has studied the distinction between the controversiality, toxicity, and resilience to toxic responses of social media posts, or whether linguistic framing can reduce toxicity in conversations of controversial topics.  
This paper answers the following research questions:
\begin{itemize}[label={}]
    \item \textbf{Q1:} What is the relationship between the controversiality of a post, its toxicity, and its resilience to toxic responses?
    \item \textbf{Q2:} Which controversial topics are resilient to toxic responses?
    \item \textbf{Q3:} How do constructive and destructive conflicts differ in terms of tone?
\end{itemize}

In exploring these questions, we make the following contributions: 
\begin{itemize}
    \item  First, we develop accurate prediction models for controversiality and toxicity resilience of social media posts. 
    We apply the models to posts in diverse Reddit communities and quantify the relationship between these two attributes. We also quantify the prevalence of toxic posts in these communities.
    \item Second, we examine the topics of posts with sparks of \emph{constructive conflict} --- posts that are resilient to toxicity despite being highly controversial.
    \item Last, we characterize the difference between linguistic features in constructive and destructive conflicts, characterized by high and low resilience to toxic responses, respectively.
\end{itemize}

We find that while a significant portion of Reddit posts attract toxic responses, response toxicity is only partially predicted by the toxicity of a post itself.
Controversial posts are often political and associate with more toxic responses. However, some political posts maintain resilience to toxicity despite being highly controversial. We demonstrate that such posts use more politeness linguistic strategies. These findings highlight the possibility of fostering constructive conflict on divisive issues.

We begin with a discussion of previous research in computational methods to measure toxicity resilience and controversiality. We then present the methods used to train our models and describe the findings. We conclude with implications for designing algorithmic curation and interfaces that foster civil discourse and reduce affective polarization.

\section{Related Work}

Ample work exists on developing and evaluating toxicity detection algorithms in online content \cite{risch2020toxic, pavlopoulos2020toxicity, sheth2022defining}, with widely used APIs such as Google's Perspective API \cite{lees2022new} and OpenAI's omni-moderation \cite{OpenAI_2024}. In this paper, we adopt a widely used definition for online and political toxicity: toxic content includes expressions of disrespect that use insulting language, profanity, name-calling, personal attacks, and the use of racist, sexist, or xenophobic terms \cite{coe2014onlineuncivil,kim2021distorting, lees2022new}. We focus on toxicity triggers, the starting points at which conversations turn toxic. We employ a similar approach to previous work, identifying these triggers as posts likely to provoke toxicity in subsequent threads \cite{almerekhi2019toxicitytriggers}.

Many studies quantify the controversiality in various ways.
Network-based approaches define controversial topics as highly clustered graphs, reflecting polarized viewpoints among social media communities~\cite{garimella2018quantifying}. Other work focuses on interactions such as edit histories on Wikipedia pages to capture the contentiousness of topics~\cite{sumi2011edit, vuong2008ranking}. Controversiality has also been inferred by linked Wikipedia pages ~\cite{dori2016controversy}.
Hybrid methods combine user-based, graph-based, and textual features to improve detection accuracy~\cite{koncar2021analysis, benslimane2021controversy}. For this study, we adopt a prior definition of controversiality as the perception of whether an issue is likely to evoke disagreement  \cite{sznajder2019controversy}. 

Prior research examines the role of linguistic features such as politeness cues in prosocial \cite{bao2021alrightconversations} or asocial \cite{zhang2018awryconversations} conversations.
Connective language ---language that signals openness to differing views--- also helps with constructive discussions
\cite{lukito2024comparing}.
This line of research has motivated the development of many tools to improve the constructiveness of online discussions. Examples include designing mobile apps to encourage conversations among people differing political views \cite{doris2013political}, and to deescalate heated threads in real-time \cite{chang2022thread}.

\section{Methodology}
\sectionname{sec:methods}{Methodology}

We capture constructive conflicts by operationalizing three properties of posts: \emph{controversiality} (C), \emph{toxicity} (T), and \emph{toxicity attraction} (TA). Note that TA is simply the opposite of toxicity resilience; in the remainder of the paper, we use TA for convenience.   
To quantify C and TA in social media posts, we develop two distinct models: the \emph{C model} and the \emph{TA model}. Both models leverage DistilBERT, a lightweight, effective, and widely used model for language representation \cite{sanh2019distilbert}.
For training and evaluation, we use two data sources. The C model is trained on Wikipedia data, while the TA model is trained on Reddit data. Both models are evaluated using Reddit data.
In the following sections, we detail the datasets, model training processes, and evaluation methods for each model.

\subsection{Toxicity Attraction}

The Reddit dataset consists of submissions (equivalent to original posts on other platforms and hereafter referred to as \textit{posts}) and comments from 50 subreddits collected using the Pushshift API\footnote{\url{pushshift.io}} between January 1 2021--December 31, 2023. These subreddits are chosen from the 150 most active subreddits\footnote{\url{reddit.com/best/communities/1}} to capture a broad range of topics with mostly text content, including political (e.g., r/Conservative, r/politics), religious (e.g., r/atheism), and general interest subreddits (e.g., r/Music, r/NoStupidQuestions, r/gaming, r/relationships, r/RandomThoughts, r/Showerthoughts). We aim to analyze conversations that have enough interactions. To do this, we extract only submissions with at least five comments. We further standardize conversation length by randomly sampling up to ten direct comments per submission (i.e., excluding replies to comments).
Next, we remove posts (submissions and comments) that were deleted by authors or moderators. We find these posts either by matching them with the labels ``[removed]'' or ``[deleted],'' or by identifying standardized moderator messages. These messages contain the keywords ``removed'' or ``deleted'' and occur more than 20 times within a subreddit. We further preprocess the text in each post by removing URLs, multiple white spaces, and HTML-specific characters, such as ``\&gt;.''
At the end of this process, the dataset contains 1,441,907 submissions and 12,866,675 associated comments, where each submission has between five and ten comments. See \appx{} for the complete list of subreddits and the number of submissions in each of them. 

The toxicity of posts can be measured using various models. We compare multiple models on a manually annotated subset of the Reddit data. The OpenAI omni-moderation-2024-09-26 model \cite{OpenAI_2024} was used throughout this work to determine toxicity scores, as it is highly accurate, achieving a ROC-AUC score of 0.91 on a set of manually annotated Reddit posts (see \appx{} for details on model selection).

\begin{figure}[ht]
\centering
\includegraphics[width=0.75\linewidth]{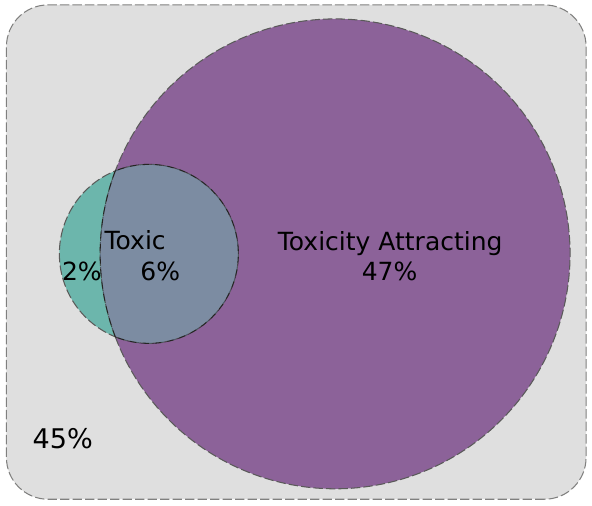}
\caption{Overlap between toxic and toxicity-attracting submissions. 45\% of all the submission are neither toxic nor toxicity attracting.}
\label{fig:ta_t_venn}
\end{figure}

We begin by exploring the relationship between toxicity and toxicity attraction. For this purpose,
a submission or comment is classified as toxic when its toxicity score exceeds 0.5 --- this threshold was chosen to balance precision and recall (see \appx{} for more details on threshold selection).
A submission is considered to be toxicity attracting if it has at least one toxic comment.
Around 8\% of submissions and 12\% of comments exhibit toxicity.
A significant portion of submissions, 53\%, are toxicity-attracting.
Only 6\% of all submissions are both toxic and toxicity-attracting. This relationship is illustrated in Fig.~\ref{fig:ta_t_venn}, highlighting the distinction between a post's toxicity and its potential to attract toxic responses. 
The distinction between toxic and toxicity attracting posts remains when different thresholds are used to define toxicity attracting, although the absolute overlap varies, as expected (see \appx{}).
From here on, we abandon any threshold‐based labels altogether: TA score is treated as a continuous measure --- the average toxicity score of all comments a submission receives—and no further binarization or cut-offs are applied.

\begin{figure}[ht]
\centering
\includegraphics[width=\linewidth]{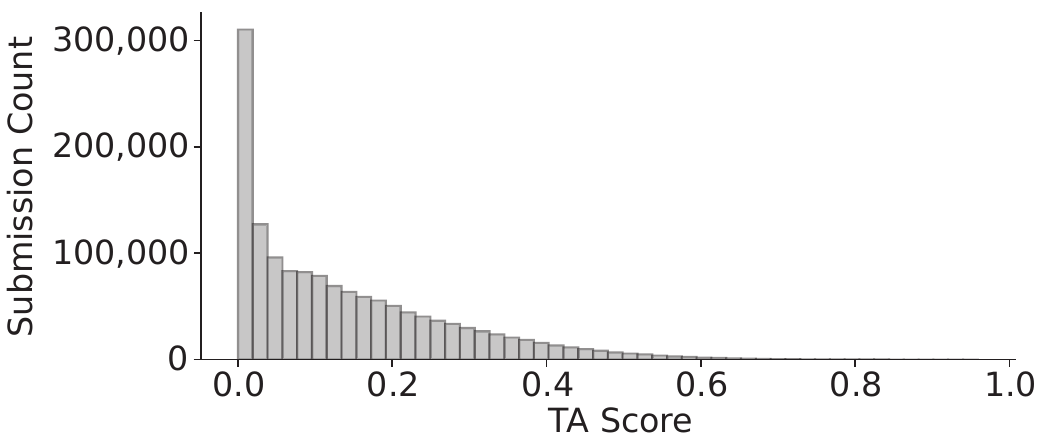}
\caption{Distribution of toxicity attraction scores.}
\label{fig:sampling}
\end{figure}

Our aim is to develop a TA regression model that takes the text of a submission as input and predicts its TA score, i.e., the average toxicity score of the comments it elicits. 
The TA model outputs a score between zero (not toxicity-attracting) and one (highly toxicity-attracting). 
It is trained using text merged from the title and body of Reddit submissions. 
We split the dataset into 70-15-15\% for training, validation and testing, respectively. Fig.~\ref{fig:sampling} shows the distribution of TA scores.

We use a pre-trained DistilBERT model, and adapt it for the TA regression task by adding a single node with linear activation function as the last layer.
We apply a grid search for hyperparameter tuning, exploring learning rates ($2 \times 10^{-5}$, $10^{-4}$) and weight decay values (0, 0.05).
The model is fine-tuned for a maximum of 5 epochs with a batch size of 32 and early stopping based on root mean squared error (RMSE) with a patience of 2 epochs.
Fine-tuning takes about 7.5 hours on a single A100X NVIDIA GPU. 
The top performing model is optimized with a learning rate of $2 \times 10^{-5}$ and a weight decay of 0. This model achieves a Spearman correlation of 0.70 on test set (0.74 in validation, both statistics $p<0.01$) with a RMSE of .10 (.09).
 
\begin{figure}
\centering
\includegraphics[width=1\linewidth]{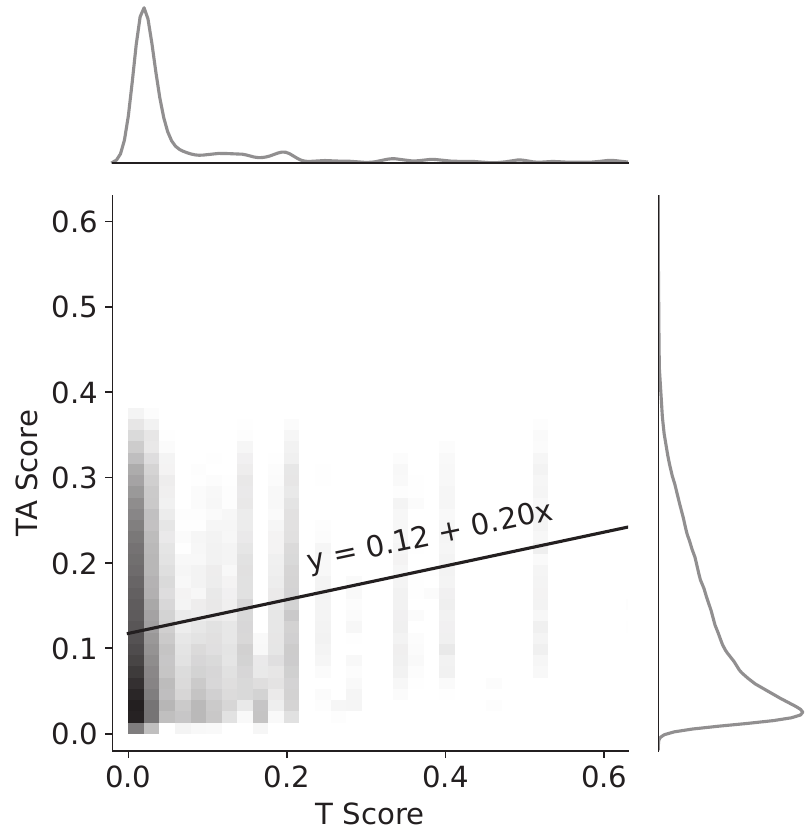}
\caption{Distribution of submission toxicity and toxicity attraction scores. Color intensity in bins reflects log‑transformed, normalized counts. The solid line represents the baseline regression model, with a coefficient standard error of $0.01$.
}
\label{fig:toxicity_vs_ta}
\end{figure}

As a baseline to evaluate our model's accuracy, we also evaluate a linear regression TA model (ordinary least squares) where the independent variable is the submission toxicity score.
This baseline achieves a RMSE of 0.13 and a Spearman correlation coefficient of 0.26 ($p < 0.01$), demonstrating that toxicity alone is a weak predictor of toxicity attraction. This relationship is further visualized in Fig.~\ref{fig:toxicity_vs_ta}.

\subsection{Controversiality}

Our controversiality model is based on both a Wikipedia dataset and a definition of controversy by \citeauthor{sznajder2019controversy}~(\citeyear{sznajder2019controversy}).
The dataset consists of 3,561 Wikipedia topics along with their short descriptions.\footnote{The dataset is released under CC-BY-SA and is available at \url{research.ibm.com/haifa/dept/vst/debating_data.shtml}}
At least ten annotators were asked to rate each topic on whether people would likely argue about it, achieving a high Cohen's Kappa agreement of 0.53.
We remove one entry missing a description and four entries with nonsensical descriptions, resulting in a total of 3,556 topics.

Examples of non-controversial topics are ``Snow leopard,'' ``copyright,'' and ``theater''; controversial topics are ``Israel, Palestine, and the United Nations,'' ``Feminism,'' and ``Pornography''; and topics with intermediate controversiality scores are ``Michael Jackson's health and appearance,'' ``Jehovah,'' and ``1956 Winter Olympics."

The C model is trained using the title and description of Wikipedia entries. We split the dataset into 70-15-15\% for training, validation and testing, respectively. We employ a pre-trained DistilBERT model, and adapt it for the C regression task by adding a single node with linear activation function as the last layer.
We apply a grid search for hyperparameter tuning, exploring learning rates ($2 \times 10^{-5}$, $10^{-5}$, $10^{-4}$) and weight decay values (0, 0.01, 0.05).
The model is trained for 10 epochs with a batch size of 16 and early stopping based on RMSE with a patience of 3 epochs. The model is optimized using RMSE loss, but the selection of the best model uses Spearman correlation as we focus on preserving relative order rather than absolute values.
The top-performing model is optimized with a learning rate of $2 \times 10^{-5}$ and a weight decay of 0.05. Training takes around ten minutes on a single A100X NVIDIA GPU. This model achieves a Spearman correlation of 0.77 ($p<0.01$) and a RMSE of 0.17 on the validation set, and a Spearman correlation of 0.76 ($p<0.01$) with a RMSE of 0.16 on the test set. This demonstrates high accuracy in identifying controversial Wikipedia topics. 

Once trained, the C model takes the text of a social media post and outputs a controversiality score between zero (non-controversial) and one (highly controversial). We assess the performance of the model by comparing the results to a manually annotated subset of the Reddit dataset. The C scores are skewed, with most posts being non-controversial. To make sure the annotation set encompasses a broad range of C scores, we sample ten posts from each decile. Two authors independently annotate comments in the sample in two rounds using the instruction: ``Mark 1 if this comment mentions a topic that people are likely to argue about, 0 otherwise.''
After the first round of independently labeling a set of 200 posts, the authors collectively review their judgments to resolve disagreements and refine the definition of controversy. 
In the second round, they independently annotate sets of 100 posts, with 50 posts overlapping among the sets to assess inter-rater agreement. This results in a Cohen's Kappa of 0.82, indicating strong agreement. Finally, the model predictions are evaluated using a labeled set of 250 posts created from both rounds, achieving an ROC-AUC of 0.90. This demonstrates high accuracy in identifying controversial content on social media.

\begin{figure}
\includegraphics[width=\linewidth]{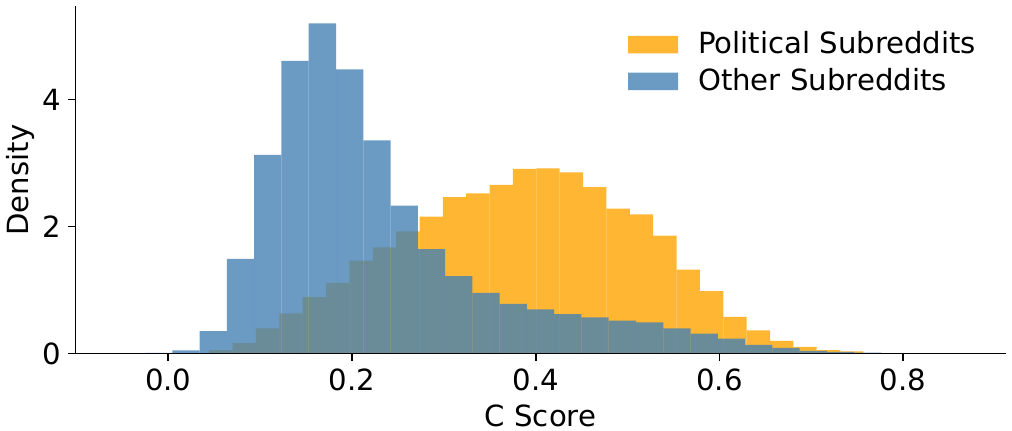}
\caption{Distribution of C scores among submissions in political and other subreddits.}
\label{fig:c_score_dist}
\end{figure}

Although a submission’s controversiality is often associated with its topic, the C model captures how controversiality manifests beyond topical content alone. Figure \ref{fig:c_score_dist} presents the distribution of C scores within political and other subreddits, showing notable variation. This indicates that our model does not merely reflect topics (which would result in minimal variation, around 0), but instead identifies nuanced signals of controversy. \appx{} provides examples to further illustrate how wording differences can influence whether a post provokes controversy \cite{salminen2020topic}.

\subsection{Linguistic Features}
\sectionname{sec:linguistic}{Quantifying Linguistic Features}

Let us outline a number of linguistic features that may play a role in the toxicity attraction of submissions. 

\begin{itemize}
    \item 
\textit{Question-asking}. 
Asking questions in online discussions has been shown to enhance likeability \cite{huang2017doesn} and perceived trustworthiness \cite{saltz2024re}. 
We represent the question-asking feature as the ratio of sentences with a question mark to all sentences in a post.

\item
\textit{Elaboration}. 
Prior work suggests that details and coherence improve the perceived quality of communication \cite{crossley2016say}. Incorporating rhetorical devices and supporting evidence --- such as narratives and factual content --- can further improve persuasiveness and acceptance \cite{feng2023effects}. It is therefore reasonable to assume that authors of longer, more elaborated texts aim to increase acceptance, making such posts less likely to elicit toxicity.
We base our definition of \textit{elaboration} on lexical diversity \cite{johansson2008lexical}, captured by the number of unique words within key lexical groups (i.e., nouns, verbs, adjectives, and adverbs) in a post. The groups are identified using part-of-speech tagging via the NLTK library \cite{bird2009natural}. 
Alternative operationalizations of elaboration in posts produce qualitatively comparable results (see \appx{}).

\item
\textit{Hedge usage}.
Hedging words/phrases such as ``perhaps,'' ``it seems,'' and ``I believe'' can be used in contentious discussions to soften assertions and reduce vulnerability to criticism \cite{lakoff1973hedges,crystal1988keeping}.
As a rhetorical strategy, hedging can enhance the persuasiveness and reception of written communication \cite{rezanejad2015cross}. While excessive hedging may undermine perceptions of authoritativeness, moderate use has been shown to improve appearances of sociability, such as warmth and approachability \cite{hosman1989evaluative}. Prior research shows that conversations starting with hedged remarks maintain civility longer than those beginning with forceful questions or direct language \cite{zhang2018awryconversations}. These findings suggest that strategic hedging may help reduce the likelihood of eliciting toxic responses. In our analysis, we quantify \textit{hedge usage} as the ratio of hedge-signaling words to the total word count of a post \cite{islam2020lexicon}.

\item
\textit{Gratitude}.
Expressions of gratitude have been shown to strengthen social connections \cite{algoe2012find}, foster reciprocal kindness \cite{mccullough2001gratitude}, and signal prosocial intentions \cite{you2006gratitude}. 
% We hypothesize that posts showing gratitude are more resilient to toxicity. 
\textit{Gratitude} is a variable representing the ratio of gratitude words to all words in a post. To identify such words, we expand and use a lexicon of gratitude words, such as ``thank you,'' and ``grateful for'' \cite{bao2021alrightconversations}. See \appx{} for the full lexicon.

\item
\textit{Name-calling}.
Name-calling directed at individuals or groups is often used as a tactic to belittle or harass others \cite{lenhart2016online, Duggan2017namecalling}. Such harsh or demeaning language has been linked to online incivility \cite{coe2014onlineuncivil}. We expect this association to hold in our dataset --- specifically, that increased usage of proper nouns is associated with higher elicited toxicity. 
To capture this, we define the \textit{name-calling} feature based on part-of-speech tagging via the NLTK library \cite{bird2009natural}, specifically the ratio of proper nouns (``NNP'' and ``NNPS'') to the total number of words in a post.

\item
\textit{Polarity}. 
Polarity measures the negative or positive emotions of a post.
The polarity of posts can have contagion effects --- negative sentiment often amplifies adversarial reactions in already polarized discussions \cite{chang2023feedback}, while positive sentiment tends to foster more positivity \cite{ferrara2015measuring, kramer2014experimental, coviello2014detecting}.
We measure the \textit{polarity} of submissions using the scores generated by VADER, a widely adopted lexicon-based sentiment analyzer 
\cite{hutto2014vader}.
\end{itemize}
% Though distinct from toxicity, polarity and toxicity frequently correlate across platforms and different models~\cite{brassard2019subversive}; in our dataset, the spearman correlation between submission toxicity and polarity is $\rho=-0.11$ ($p<0.01$).

\section{Results}
\sectionname{sec:results}{Results}

\subsection{Controversiality and Toxicity Resilience}

We study the relationship between the controversiality of posts and their resilience to toxic responses using the predicted C and TA scores.
Aligning with our intuition, controversial posts are more likely to draw toxic responses. The Spearman correlation between TA and C values is 0.48 ($p<0.001$). This suggests that controversiality is a stronger predictor for the toxicity attraction of a post than its toxicity. 

\subsection{Topics}

\begin{figure*}
\centering
\includegraphics[width=\linewidth]{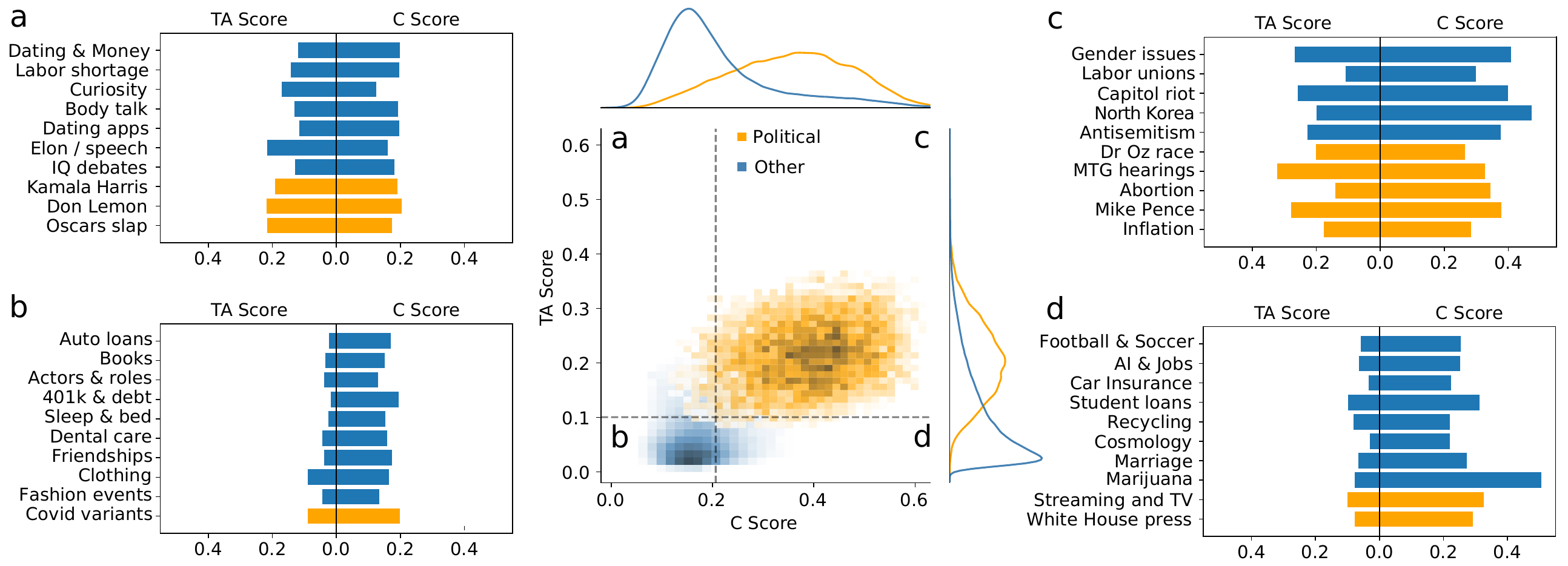}
\caption{
The relationship between toxicity attraction (TA) and controversiality (C) in Reddit. Middle panel: Distribution of TA and C scores of submissions in political (orange) and non-political (blue) subreddits. The dashed lines represent median TA and C values, separating quadrants of low/high C/TA. The bar plots in panels (a)-(d) illustrate topics sampled from submissions with TA and C values in these quadrants, along with their average scores. Topics are obtained using BERTopic and are abbreviated for clarity (see \appx{} for the full keyword lists). (a) Topics sampled from submissions with high TA, low C scores. (b) Topics sampled from submissions with low TA, low C scores. (c) Topics sampled from submissions with high TA, high C scores. (d) Topics sampled from submissions with low TA, high C scores.
}
\label{fig:ta_vs_c}
\end{figure*} 

Let us examine how controversiality and resilience to toxic responses depend on different topics. To begin, we consider political and non-political content, as illustrated in Fig.~\ref{fig:ta_vs_c} (middle panel).
Submissions from political subreddits (i.e., r/Conservative, r/Liberal, r/politics) are significantly more controversial than non-political subreddits (e.g., r/RandomThoughts, r/gaming, and r/changemyview). Median C values are significantly different for political and non-political subreddits: 0.39 and 0.20, respectively; Mood's median test statistic 16,937, $p<0.01$.
Furthermore, submissions from political subreddits are more likely to attract toxicity than non-political subreddits. The median TA values for these subreddits are significantly different: 0.22 and 0.09, respectively; Mood's median test statistic 21,399, $p<0.01$. While the prevalence of political content in non-political subreddits is less than in subreddits dedicated to political topics, the former may occasionally include political discussions.

The distribution of predicted TA and C scores for Reddit submissions and the topics associated with these scores are visualized in Fig.~\ref{fig:ta_vs_c}. 
The topics are generated using BERTopic, an effective and widely used topic modeling technique \cite{grootendorst2022bertopic}; implementation details are provided in \appx{}. We remove topics supported by less than ten submissions. Fig.~\ref{fig:ta_vs_c}a-d illustrate combinations of TA and C scores across quadrants, divided by the median TA and C scores. For each quadrant, we randomly draw six topics from political, and six from non-political subreddits. 

As expected, most low-controversy submissions are non-political (Fig.~\ref{fig:ta_vs_c}a,b). 
Among these, high TA submissions often involve personal insults or body references (Fig.~\ref{fig:ta_vs_c}a). Low TA submissions typically cover mundane topics such as books, car loans, dentists, and fashion, which are less likely to trigger strong disagreement or toxic language. Note that only one political topic appear in the low-controversy region (Fig.~\ref{fig:ta_vs_c}b).

Constructive conflicts --- submissions that are controversial (high C), yet resilient to toxicity (low TA) --- are demonstrated in Fig.~\ref{fig:ta_vs_c}d. These submissions mention policy-related topics such as student loan forgiveness, artificial intelligence, big bang, and marijuana usage.
Other issues in the high C - low TA quadrant, not shown in the figure, are climate change and alternative energy sources, healthcare and insurance policies, and vegetarianism. 
This underscores the possibility of discussing critical controversial issues in a respectful manner on social media platforms.

In contrast, toxicity-attracting submissions are about religious themes or highly polarizing political issues such as abortion rights, the January 6 riot, racial issues, and LGBTQ+ rights (Fig.~\ref{fig:ta_vs_c}c). Other topics in this quadrant (not depicted) are the Israel-Gaza conflict, the Ukraine war and school shootings.

\subsection{Linguistic Features}

The topics associated with destructive conflict --- submissions that are controversial (high C) and toxicity-attracting (high TA) --- deal with important issues that merit dialogue (Fig.~\ref{fig:ta_vs_c}c). Could conversations about these topics be reframed to become civil and constructive? 
To explore this question, we compare the linguistic cues in highly controversial submissions with low and high TA scores.

Consistent with our intuition, submissions employing civil features tend to attract less toxicity. In particular, question-asking, elaboration, hedges, gratitude, and positive polarity are negatively correlated with TA scores (Table \ref{tab:linguistic_feat}, left column). Conversely, negative linguistic cues, such as negative polarity and name-calling, are associated with higher TA scores. 
These findings align with prior research, which demonstrates that politeness cues such as question-asking and elaborated text foster more prosocial conversations \cite{bao2021alrightconversations}, while the absence of hedge and gratitude expressions are linked to conversations that begin civilly but eventually derail into incivility \cite{zhang2018awryconversations}. 

Most of these patterns hold for non-controversial submissions as well. However, name-calling notably increases toxicity attraction in controversial submissions, while having no significant effect on non-controversial submissions. A possible explanation is that referencing individual names in controversial discussions may signal direct confrontation or the targeting of political figures or parties, which can heighten response toxicity. On the other hand, calling someone by name in non-controversial settings does not necessarily have a negative connotation. We also run a regression model for TA scores using C scores along with each linguistic feature as independent variables, showing that the effects of most features are robust (see \appx{}).

\input{tables/linguistic_features_statistics.tex}

Let us further illustrate the role of linguistic features in evoking toxicity with examples from our dataset.
We select a set of submissions discussing highly polarized issues according to a recent Gallup Poll \cite{Newport_2024}. Table \ref{tab:linguistic_ta_c_scores} shows pairs of submissions about the same high-C topics, but markedly different TA scores for comparison.

For the topic of abortion, a question-seeking post that includes a hedging symbol (``think'') has a significantly lower TA score (39th percentile) than a post with name-calling and an accusatory tone (TA in the 99th percentile).
For the topic of healthcare, a factual question about policy outcomes demonstrates how elaboration and asking questions can reduce the likelihood of attracting toxicity compared to inflammatory language targeting. These submissions have TA scores in the 10th and 75th percentiles, respectively.
For the topic of gun laws, the same features result in TA scores in the 30th and 95th percentiles, respectively.
The effect of showing gratitude and hedged phrasing is seen across examples on different topics, where it results in lower TA scores for submissions about LGBTQ+ and climate change (TA scores in the 33th and 6th percentiles, respectively). On the contrary, dismissive rhetoric increases TA (scores in the 99th and 92th percentiles for submissions about LGBTQ+ and climate change, respectively).
These examples show that highly polarizing topics can be discussed civilly using various linguistic features.

\input{tables/language_examples}

\section{Discussion}
\sectionname{sec:discussion}{Discussion}

The rise of affective polarization and its link to social media usage \cite{lorenz2023systematic}, has raised concerns about the potential harm caused by algorithms employed by social media platforms. These concerns have spurred research into alternative, ``bridging'' algorithms that promote cross-ideological understanding and respect \cite{ovadya2023bridging, piccardi2024reranking}. 
One way to operationalize bridging heuristics is by encouraging diverse exposure to cross-cutting content \cite{gutmann2009democracy, bohman1997deliberative, levy2021social}. We propose that constructive conflicts --- posts that are controversial, yet resilient to toxicity --- can provide a foundational condition to promote such exposure.

We develop accurate machine learning models to quantify the controversiality and toxicity resilience of posts, and analyze the results across diverse Reddit communities. Previous research shows that across platforms like Facebook, Gab, and Twitter (but not Reddit), controversy is positively correlated with increased toxicity \cite{avalle2024persistent}.
Our work bridges the literature on online incivility and conversational framing, showing that political controversy and toxicity are not inherently linked, and that language framing plays a critical mediating role. This suggests the feasibility of achieving a constructive online environment. 

The proposed attributes can be flexibly integrated into recommendation algorithms to align with specific goals. For instance, users in open ecosystems like Bluesky could use these metrics to create custom feed generators (e.g., FeedGenerator\footnote{\url{github.com/bluesky-social/feed-generator}}) to prioritize posts based on toxicity resilience, controversiality, or both. 
Toxicity resilience predicts the likelihood of toxic subsequent comments to a post, accounting for posts at risk of high toxicity that might be overlooked in content-level analyses. More importantly, optimizing only for low-toxicity content may disproportionately favor neutral content and overlook controversial content that attracts diverse perspectives.
By simultaneously considering controversiality together with toxicity resilience, we can create a ranking heuristic to demote negativity while retaining a diversity of viewpoints. 

Since posts that attract toxicity are often linked to high-arousal emotions that drive online engagement \cite{ferrara2015measuring, robertson2023negativity}, demoting these posts could reduce overall platform engagement.
The most valid usability test to examine the effect of ranking heuristics on engagement would involve deploying the ranking in a field experiment \cite{piccardi2024reranking}. However, such a design is time-consuming and costly, and therefore remains outside the scope of the current work.

Observational analysis can offer insights into the potential effects of our framework on engagement, but it cannot establish causality, and may not even yield accurate correlations. 
Specifically, data available through the platform reflects the engagement and exposure shaped by existing recommendation algorithms, while posts surfaced through any alternative rankings would likely generate different engagement dynamics. Consequently, simply correlating proposed ranking scores with engagement metrics from Reddit's current system likely overestimates or misrepresents the true impact of our framework. Further research is needed to fully understand and address this trade-off before integrating the heuristic into platform ranking algorithms.

The proposed methods enable the early identification of discussions at risk of escalating toxicity. Platforms and moderators can leverage these predictions to implement safeguards, such as adding warnings or context labels to posts that are likely to elicit toxic responses.

Our analysis reveals that even conversations on deeply divisive issues like abortion, gun laws, and LGBTQ+ rights can resist toxic responses through strategic language choices. Specific linguistic strategies, such as questioning, hedging, and expressing gratitude, increase a post's resistance to toxic responses. These findings inform the design of platform affordances that encourage healthier online discussions. For instance, prompts could guide users toward interactions that reduce the risk of toxicity. Another possible application of our analysis is in platform interfaces that highlight politeness cues when displaying comments.

While powerful large language models could be useful aids to algorithmic ranking \cite{piccardi2024social} and rephrasing issues \cite{costello2024durably}, their high computational cost can cause delay in real-time user experience, decreasing user engagement.
The proposed models utilize DistilBERT, a lightweight transformer model with short inference time that enables efficient real-time integration into recommendation systems. Research indicates that BERT-based architectures outperform GPT models in detecting certain linguistic features \cite{lukito2024comparing}, supporting our architectural choice.

This study has several limitations. First, we rely on the OpenAI omni-moderation model to identify toxic submissions and comments, which means our findings inherit any shortcomings or biases of the model. Second, the proposed toxicity attraction model is trained using Reddit data and our specific definition of toxicity, limiting their generalizability to other platforms.
We operationalize the toxicity attraction of a post as the average toxicity of its comments. Our results are robust under an alternative definition, where the TA score is the ratio of a submission's toxic comments.
This alternative method strongly correlates with our proposed TA metric (Spearman correlation: $r=0.92, p<0.01$). In addition, despite their high accuracy, our proposed models are not immune to prediction errors, such as assigning a high C score to low-controversial posts or a high TA score to posts that are not truly toxicity-attracting. Applications utilizing these models should exercise caution by fine-tuning them with platform- and use case-specific data and regularly retraining them with updated datasets and annotations to maintain reliability.
Lastly, implementing the controversiality model in real-world applications would require a frequently-updated method for generating predictions. 
However, our model shows strong potential in capturing recent topics effectively. For example, despite being trained on 2019 Wikipedia data, it successfully identifies events from subsequent years, such as the January 6 riot, as highly controversial topics, indicating reasonable robustness.

\section{Ethics Statement and Broader Impact}
\sectionname{sec:ethics}{Ethics Statement and Broader Impact}

We have taken several steps for ethical considerations. First, we do not use any personally identifiable information (PII).
We utilized two datasets for our research. The first is the Wikipedia Dataset by \cite{sznajder2019controversy}, which contains topics and their short descriptions. This dataset does not contain any PII.
The second is the Reddit dataset, consisting of public submissions and comments. While the posts may include PII, such as usernames or self-disclosures, the dataset is limited to publicly available posts. We collected this dataset using the Pushshift API, adhering to the platform's terms of service to ensure compliance.
In addition, the dataset is stored on a server with restricted access. Lastly, we release the data in Zenodo\footnote{\url{zenodo.org/records/17167317}} and the source code in a GitHub repository \footnote{\url{github.com/ocseckin/constructive_conflict}}.
To uphold ethical standards, the curated datasets exclude raw text and any PII, providing only post IDs and extracted features, such as polarity, toxicity scores, or indicators like the presence of a question mark. These features cannot be reverse-engineered to reveal PII.

Amid increasing concerns of polarization, our study enhances the understanding of bridging conversations by providing insights on how conflicts are discussed online. We hope to inform the design of social media algorithms and interfaces that would promote dialogues productive to democratic deliberation. 

\section{Acknowledgements}

We are grateful to Giovanni Luca Ciampaglia, Do Won Kim, and Saumya Bhadani for helpful discussions. This work was supported in part by the Swiss National Science Foundation (Sinergia grant CRSII5\_209250) and by the Knight Foundation. This work used the IU JetStream 2 computational infrastructure through allocation CIS240118 from the Advanced Cyberinfrastructure Coordination Ecosystem: Services \& Support (ACCESS) program, which is supported by National Science Foundation grants \#2138259, \#2138286, \#2138307, \#2137603, and \#2138296 \cite{hancock2021jetstream2, boerner2023access}.

\bibliography{aaai25}

\appendix
\section{Appendices}

\subsection{Dataset} 

Subreddits and number of submissions: NoStupidQuestions (280,938), antiwork (99,514), politics (95,296), worldnews (94,272), WhitePeopleTwitter (69,887), gaming (62,756), AskUK (58,110), Conservative (57,217), Tinder (45,358), relationship\_advice (43,314), movies (42,698), mildlyinfuriating (41,124), dating (33,207), mildlyinteresting (30,440), news (26,512), aww (24,391), RandomThoughts (23,174), atheism (20,440), personalfinance (19,448), popculturechat (19,182), AskAnAmerican (18,297), Bitcoin (17,113), Showerthoughts (16,302), Music (14,674), technology (14,063), relationships (13,818), nottheonion (12,852), CasualConversation (11,576), Frugal (11,206), socialskills (10,781), selfimprovement (10,351), changemyview (10,242), Jokes (10,159), 4chan (9,965), WTF (9,667), RoastMe (8,784), tifu (7,876), science (7,231), introvert (7,041), 3amjokes (5,723), getdisciplined (4,546), MaliciousCompliance (4,011), HistoryPorn (3,463), sports (3,348), UpliftingNews (3,203), nosleep (2,877), GetMotivated (2,497), Liberal (1,493), AskEconomics (773), OutOfTheLoop (697)

\subsection{Choosing the Toxicity Detection Model}

We choose the best toxicity detection model by comparing the model predictions with manual annotations. Three state-of-the-art models are considered: (1) toxic-BERT \cite{Detoxify}, (2) Perspective API \cite{lees2022new}, and (3) OpenAI omni-moderation-2024-09-26 \cite{OpenAI_2024}, the latest harmful content detection model provided by \citeauthor{OpenAI_2024} as of October 2024. 

These models assign scores to multiple categories. The toxicity score inferred by each model is the maximum among the categories. We compare the inferred toxicity scores against our ground truth annotations to assess model performance. To align with our definition of toxicity, we exclude the ``self-harm/intent'' and ``violence'' categories from the omni-moderation-2024-09-26 model categories.\footnote{\url{platform.openai.com/docs/guides/moderation}} Similarly, ``obscene'' and ``profanity'' are not considered among the categories of toxic-BERT\footnote{\url{huggingface.co/unitary/toxic-bert}} and Perspective API.\footnote{\url{developers.perspectiveapi.com/s/about-the-api-attributes-and-languages}} 

% To find candidate toxic posts, we employ a toxicity detection model based on BERT, toxic-BERT \cite{Detoxify}, which has demonstrated robust performance across diverse datasets. This model assigns a toxicity score $t$ to each post in our Reddit dataset. We use these scores to construct balanced samples for manual annotation. Half of each annotation set includes non-toxic posts ($t<0.5$) and another half are toxic posts ($t>0.5$). This strategy ensures that we evaluate the model’s performance across a broad range of toxicity levels.

Annotators are asked to label a post as toxic or non-toxic. Since one of our models is Perspective API, we provide Google's Perspective API\footnote{\url{perspectiveapi.com/how-it-works/}} definition of toxicity to annotators, namely, ``a rude, disrespectful, or unreasonable comment that is likely to make people leave a discussion.''  
% Since these definitions guide human annotation, the dataset labels vary accordingly, influencing the predictions of models trained on these datasets \cite{fortuna2020toxic}.
Two annotators participate in the annotation. In the first round, they independently annotated a set of 100 posts. After this, they compare judgments, followed by a discussion to resolve discrepancies and refine the definition of toxicity. 
The second round annotation set contains 250 comments, where 150 comments are annotated independently, and 50 comments in common. We calculate the inter-rater agreement using the 50 overlapping comments, achieving a Cohen’s Kappa of 0.81, indicating strong agreement.
Finally, all the annotated labels, totaling 350 comments, are used to evaluate the performance of the toxicity detection models. The best performance is by omni-moderation-2024-09-26 model, with a ROC-AUC score of 0.91. Perspective API and toxicBERT achieves ROC-AUC scores of 0.87 and 0.84, respectively.

\subsection{Choosing the Threshold for Toxicity Model}
\begin{figure}[h]
\centering
\includegraphics[width=1\linewidth]{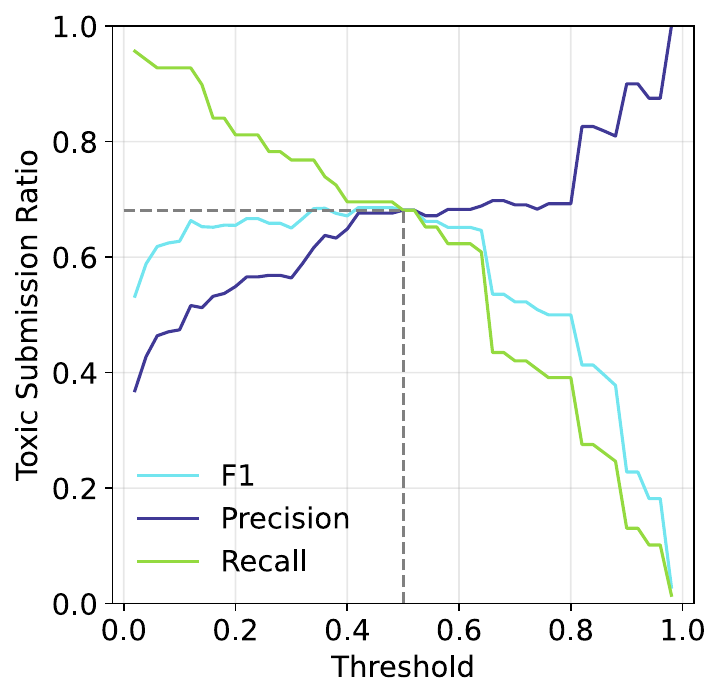}
\caption{F1, Precision and Recall scores based on different thresholds for toxicity score.
}
\label{fig:toxicity_threshold}
\end{figure}

To report the overlap between toxic and toxicity attracting submissions, we determine the optimal threshold for toxicity scores using our manually annotated dataset. Among thresholds ranging from 0 to 1, we find that 0.5 achieves the highest performance, yielding an F1-score of 0.68 (see Fig.\ref{fig:toxicity_threshold}).

\subsection{Overlap of Toxic and Toxicity Attracting Posts}

\begin{figure}
\centering
\includegraphics[width=1\linewidth]{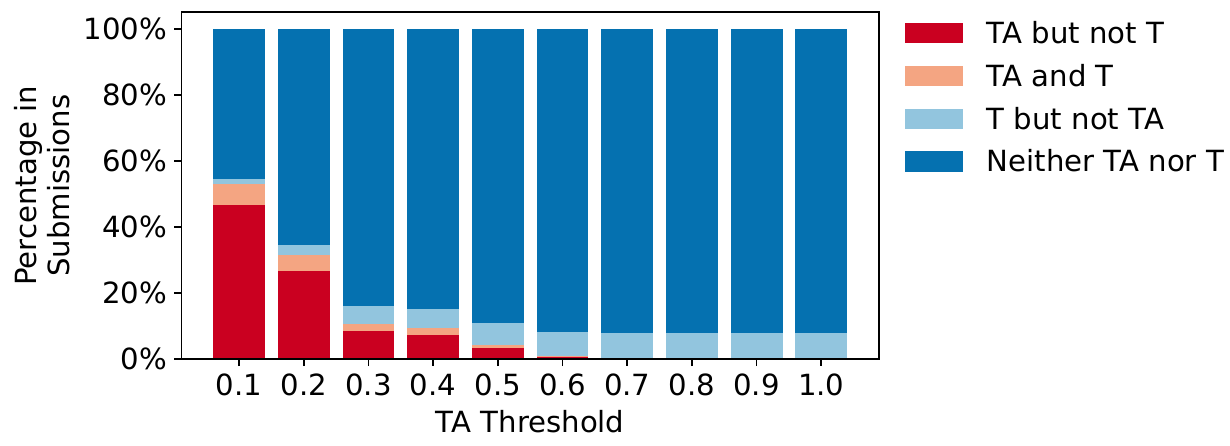}
\caption{The proportion and overlap of toxic (T) and toxicity-attracting (TA) submissions across varying toxicity-attraction thresholds.}
\label{fig:ta_threshold}
\end{figure}

Fig.\ref{fig:ta_threshold} illustrates how varying the toxicity-attraction (TA) threshold affects the overlap between toxic (T) and toxicity-attracting (TA) submissions. The percentage of toxicity-attracting but not toxic posts for thresholds 0.2, 0.3, 0.4, and 0.5 are 27\%, 8\%, 7\%, and 3\%, respectively. The percentage of toxicity-attracting and toxic posts for the same thresholds are 5\%, 2\%, 2\%, and 1\%, respectively. The percentage of toxic but not toxicity-attracting posts are 3\%, 6\%, 6\%, and 7\%, respectively.

\subsection{Within-topic Variation of C Scores}

Table \ref{tab:posts_w_diff_c} provides examples of posts discussing the same issue but receiving different C scores. For instance, two submissions about immigration reform in Germany are scored 0.67 and 0.25, indicating that how a topic is framed or expressed plays a role in shaping its C score.

\input{tables/within_topic_c_scores}

\subsection{Extracting Topics with BERTopic}
We use BERTopic to infer topics reflected in submissions. To do so, we first create a document-term matrix where we ignore terms that occur strictly lower than 10 times in the dataset. We then pass this matrix to BERTopic with number of topics set to ``automatic'' (nr\_topics = ``auto''), which automatically reduces the number of topics using HDBSCAN \citep{mcinnes2017hdbscan}. 
This method clusters the submissions into coherent topical groups and produces, for each cluster, a list of four representative keywords.
In our analyses, we do not report any topic clusters supported by fewer than ten submissions. We shortened the keyword lists so the labels would fit cleanly in figures and tables.

\subsubsection{Keyword Lists for Fig.~\ref{fig:ta_vs_c}.}
Dating \& money (``she, gf, her, money''); Labor shortage (``labor, shortage, workers, nobody''); Curiosity (``wonder, interesting, why, wow''); Body talk (``boobs, breasts, nipples, breast''); Dating apps (``apps, dating, app, bumble''); Elon / speech (``elon, centrist, speech, conspiracies''); IQ debates (``iq, intelligence, smart, smarter''); Kamala Harris (``kamala, harris, biden, vp''); Don Lemon (``don, lemon, cnn, cnns''); Oscars slap (``smith, oscars, chris, slap''); Auto loans (``car, loan, vehicle, miles''); Books (``books, book, read, library''); Actors \& roles (``actor, actors, role, roles''); 401k \& debt (``401k, mortgage, roth, debt''); Sleep (``sleep, wake, asleep, bed''); Dental care (``teeth, dentist, tooth, dental''); Friendships (``friends, friendships, group, social''); Clothing (``wear, dress, wearing, pants''); Fashion events (``fashion, highlight, gala, archives''); COVID variant (``variant, cdc, infection, mild''); Gender issues (``gender, trans, gay, nonbinary''); Labor unions (``union, unions, strike, amazon''); Capitol riot (``capitol, police, riot, officers''); North Korea (``korea, north, kim, missile''); Antisemitism (``hitler, antisemitism, nazi, holocaust''); Dr Oz race (``oz, dr, pennsylvania, john''); MTG hearings (``greene, taylor, rep, committee''); Abortion (``abortion, roe, abortions, wade''); Mike Pence (``pence, mike, trump, capitol''); Inflation (``inflation, biden, joe, bidens''); Football \& soccer (``football, sports, soccer, sport''); AI \& jobs (``ai, artificial, automation, jobs''); Car insurance (``insurance, car, coverage, damage''); Student loans (``student, loan, loans, forgiveness''); Recycling (``trash, bin, plastic, recycling''); Cosmology (``universe, infinite, bang, expanding''); Marriage (``married, marriage, divorce, marry''); Marijuana (``marijuana, cannabis, weed, recreational''); Netflix (``netflix, streaming, subscription, tv''); White House press (``press, secretary, psaki, jen'').

\subsection{Linguistic Features}

\subsubsection{Gratitude Lexicon.}\label{appx:gratitude}

thanks, contented, blessed, thank you, thankful for, grateful for, greatful for, my gratitude, i appreciate, made me smile, make me smile, i super appreciate, i deeply appreciate, i really appreciate, bless your soul, made my day, tysm, thx, shout out to.

\subsubsection{Elaboration.}
We evaluated three operationalizations of elaboration: (1) total token count, (2) lexical-word count – number of content words (nouns, verbs, adjectives, adverbs), (3) Measure of Textual Lexical Diversity (MTLD) – a lexical-diversity metric in which higher scores reflect a wider range of vocabulary rather than repetitive wording, signaling the expression of a broader set of ideas \citep{mccarthy2010mtld}. 
All three metrics show highly similar patterns: each correlates significantly with TA score (all $p < 0.01$) for both controversial and non-controversial posts.
Text length has the strongest association for non-controversial submissions ($\rho = -0.32$) and a comparable effect for controversial ones ($\rho = -0.24$). Lexical-word count follows the same trend ($\rho = -0.31$ non-controversial, $\rho = -0.18$ controversial). MTLD: $\rho = -0.30$ non-controversial, $\rho = -0.24$ controversial. Lexical-item count is strongly correlated with both alternative elaboration metrics: its Spearman correlation with MTLD is  0.80, and with total token count is an even tighter 0.98, both $p<0.01$.

\subsubsection{Regression Analysis.}
We also run a ordinary least squares regression analysis with the described linguistic features as independent variables, and TA score as the dependent variable. 
Before running the regression analysis, we assess multicollinearity by examining the variance inflation factor (VIF) for each variable. Since none of the features exhibits a VIF greater than 3, we include all variables in the analysis. The results are presented in Table~\ref{appx:tab:regression}. All coefficient signs are consistent with the correlations reported in the main text, except for the text length.

\input{tables/regression}

\end{document}

%% file: tables/linguistic_features_statistics.tex
\begin{table}
\centering
\begin{tabular}{l r r r r}
\toprule
\multirow{2}{*}{Linguistic feature} & \multicolumn{2}{c}{Controversial} & \multicolumn{2}{c}{Non-controversial} \\
\cmidrule(r){2-3} \cmidrule(r){4-5}
& Corr. & \% & Corr. & \% \\
\midrule
Question asking & -0.44 & 47 & -0.38 & 50 \\ 
Elaboration & -0.18 & 100  & -0.31 & 100 \\ 
Hedge usage & -0.16 & 55 & -0.17 & 56 \\ 
Gratitude usage & -0.06 & 3 & -0.11 & 5 \\ 
Positive polarity & -0.10 & 33 & -0.13 & 46 \\
Negative polarity & 0.11 & 48 & 0.08 & 26 \\
Name calling & 0.36 & 81 & 0.002 & 64 \\ 
% Toxicity score & 0.27 & 100 & 0.23 & 100 \\ 
\bottomrule
\end{tabular}
\caption{Correlations between linguistic feature usage and TA scores for controversial and non-controversial Reddit posts, along with the percentage of posts in which these features appear.
Point-biserial correlation is used for binary features, i.e., ``Positive Polarity'' and ``Negative Polarity,'' while Spearman correlation is used for other features.
All correlations are significant with ($p<0.01$) except for name calling, which is non-significant.}
\label{tab:linguistic_feat}
\end{table}

%% file: tables/language_examples.tex
\begin{table*}
    \centering
    \resizebox{\textwidth}{!}{
    \begin{tabular}{c p{12cm} c c}
        \toprule
        \textbf{Topic} & \textbf{Submission} & \textbf{TA \%} & \textbf{C \%} \\ \midrule
        \multirow{2}{*}{Abortion} 
        & VP Kamala Harris continues to tell Christians they should support the killing of preborn humans.
        & 99 & 91 \\ \cmidrule{2-4}
        & Should abortion be limited? I'm asking if abortion should be limited around 12-15 weeks? By then the fetus has a heartbeat, nervous system, functional organs, and yes eyes and mouth. Also women have bumps by then too. 99\% of women know they are pregnant before 12 weeks and can make decision. I don't think abortion should be totally banned but 15 week is a reasonable compromise.
        & 39 & 96 \\ \midrule
        \multirow{2}{*}{Healthcare} 
        & If the US was actually serious about supporting small businesses, they would enact universal healthcare. Working for a small business is effectively a death sentence, either financially or physically or both. Corporations hold the golden phallus of health care to dangle in front of your face. No health care only strengthens corporate asphyxiation on workers.
        & 75 & 98 \\ \cmidrule{2-4}
        & Would making healthcare free in the U.S. be more expensive due to higher taxes, or cheaper due to no insurance payments/co-pays? I’ve been wondering this since the topic of “If we make healthcare free our taxes will go up”. Although this is obviously true, I find that people forget they also wont be directly shilling out money every year for their insurance. So my question is in the title, I know taxes in other countries are a lot higher than in the U.S. because of that reason, but when incorporating the money you technically save from insurance and copayments would it still be that much of a difference? & 10  & 88 \\ \midrule
        \multirow{2}{*}{Gun Laws} 
        & Enough is enough. Anyone who opposes gun control in this country is psychopathic unamerican scum. & 95 & 98 \\ \cmidrule{2-4}
        & What are actual ways where gun laws can be re-established where both parties can be satisfied (whether for or against gun-control)? & 30 & 99 \\ \midrule
        \multirow{2}{*}{LGBTQ+} 
        & Hating someone for being gay is a lifestyle choice. & 99 & 84 \\ \cmidrule{2-4}
        & what’s it called when you feel as if you were born the wrong gender but don’t wanna go through with a transition of any sort because you know you wouldn’t really be that gender? i in no way mean to invalidate anyone at all but i’ve been having some thoughts bout myself but i know it wouldn’t be the same as being born as the opposite gender, i wanna look more into this so if someone could point me in the right direction of what to be looking up that would be great, and thank you in advance :) & 33 & 51 \\ \midrule
        \multirow{2}{*}{Climate Change} 
        & This idiot still pushing climate change as a hoax. Same as he did with COVID. & 92 & 72 \\ \cmidrule{2-4}
        & If global warming becomes a problem couldn’t everyone just move to a cooler place Like say the temperatures get really hot Couldn’t everyone just move to a place/country that is colder? Or for flooding couldn’t everyone just move inland and be all good? & 6 & 81 \\ \bottomrule
    \end{tabular}
    }
    \caption{Examples of posts discussing highly polarized issues (according to a recent Gallup Poll \cite{Newport_2024}) in our dataset. Random pairs of Reddit posts about similar topics with markedly different TA scores. Percentiles for TA and C scores are shown, where higher percentiles indicate higher scores. Topics are obtained using BERTopic.  Topic names are abbreviated for clarity: Abortion (abortion\_roe\_abortions\_wade), Healthcare (insurance\_healthcare\_medical\_gp), Gun Laws (gun\_guns\_shootings\_weapons), LGBTQ+ (gender\_trans\_gay\_nonbinary), and Climate Change (climate\_warming\_global\_change).}
    \label{tab:linguistic_ta_c_scores}
\end{table*}

%% file: tables/within_topic_c_scores.tex
\begin{table*}[h]
    \centering
    \resizebox{\textwidth}{!}{
    \begin{tabular}{c p{12cm} c c}
        \toprule
        \textbf{Topic} & \textbf{Submission} & \textbf{C Score} & \textbf{C \%} \\ \midrule
        
        \multirow{2}{*}{Immigration in Germany} 
        & Germany is planning to reform its immigration policies. The U.S. should, too. & 0.67 & 99 \\ \cmidrule{2-4}
        & Germany to ``recruit workers from abroad'' to ease airport chaos & 0.25 & 63 \\ \midrule
        
        \multirow{2}{*}{Cannabis Usage} 
        & Adolescent cannabis use and later development of schizophrenia: An updated systematic review of six longitudinal studies finds ``Both high- and low-frequency marijuana usage were associated with a significantly increased risk of schizophrenia.'' & 0.83 & 99 \\ \cmidrule{2-4}
        & Can you get high of weed passively? The dentis just said, I cant smoke or drink for a week or so. Well, I cant smoke, but what if I just burn it and whiff it? & 0.15 & 23 \\ \midrule
    
        \multirow{2}{*}{Taxation} 
        & Why does sales tax exist? & 0.46 & 90 \\ \cmidrule{2-4}
        & Why is the tax not on the price tag? Ok so for reference and if this a stupid question, I’m from a country where the price tag is the final price. Tax included. But I’ve been to the US (recently I knew Canada has it too) the final price is added at the checkout. I find it kinda annoying because how do you know the exact price before checking out? Is there a way? & 0.23 & 58 \\ \midrule
        
        \multirow{2}{*}{Racing Games} 
        & What Happened to Racing Games Lately? Anyone knows why there are very less racing games nowadays and most of them are not even that enjoyable. Just a decade ago there was huge catalog of them with the likes of Need For Speed, Midnight Club, Rumble Racing, Blur, Juiced, Driver, Grid, Burnout Series. Heck even crazy taxi not even a racing game was so good for its time. The most wonderful thing was all of these had local couch co-op and were really fun to play. Always wondered if these games had a open world mode. But now years later we have only Forza. Dont get me wrong, Forza also is a very good game but misses the joy of Street racing. Need for speed are released with years of gap and have no split screen and are very short and full of dlc and not that much exciting anymore. Whats up with the racing devs lately. How the heck was most old racing games were so good and not now. & 0.18 & 38 \\ \cmidrule{2-4}
        & Looking for a new car racing game. Something inbetween a racing sim (project cars) and an arcade game (nfs). I couldn't really get into the new forza horizon, I have wreckfest but want a tarmacy game lol I'm on PC. Thanks. & 0.12 & 14 \\ \bottomrule
    \end{tabular}
    }
    \caption{Examples of posts discussing same issues with differing C scores. C scores and their percentiles are shown, where higher percentiles indicate higher scores(controversiality). Topics are obtained using BERTopic.  Topic names are abbreviated for clarity: Immigration in Germany (germany\_immigration\_immigrants\_bolster), Cannabis Usage (marijuana\_cannabis\_weed\_recreational), Racing Games (racing\_driving\_games\_game), and Taxation (tax\_price\_sales\_prices)}
    \label{tab:posts_w_diff_c}
\end{table*}

%% file: tables/regression.tex
\begin{table}[h]
\centering
\begin{tabular}{l c c}
\toprule
Variable & Coefficient & Std Err \\
\midrule
intercept*** & 0.054 & 0.001 \\
c\_score*** & 0.344 & 0.001 \\
question\_ratio*** & -0.067 & 0.001 \\
gratitude\_ratio & -0.017 & 0.022 \\
proper\_noun\_ratio*** & 0.046 & 0.001 \\
lexical\_item\_count & 1e-06 & 1e-06 \\
hedge\_ratio*** & 0.017 & 0.003 \\
polarity*** & -0.018 & 1e-06 \\
% submission\_toxicity & 0.1695 \\
%                             & (0.001) \\
\bottomrule
\end{tabular}
\caption{Ordinary least squares regression results. ***($p<.01$).}
\label{appx:tab:regression}
\end{table}